\documentclass[twocolumn,floatfix,pre,a4paper,showpacs]{revtex4-1}
\pdfoutput=1
\usepackage{graphicx}
\usepackage{amssymb}
\usepackage{amsmath}
\usepackage[utf8]{inputenc}
\usepackage{cancel}
\usepackage[usenames,dvipsnames]{color}
\usepackage{psfrag}
\usepackage{epsfig}
\usepackage{bbm}
\usepackage{bm}
\usepackage{dsfont}
\usepackage{soul}

\usepackage{colonequals}

\begin{document}
\title{Delay-induced transport in a rocking ratchet under feedback control}
\author{Sarah A.~M.~Loos, Robert Gernert, and Sabine H.~L.~Klapp}
\affiliation{
  Institut f\"ur Theoretische Physik,
  Hardenbergstr.~36,
  Technische Universit\"at Berlin,
  D-10623 Berlin,
  Germany}

\date{\today}
\begin{abstract}
Based on the Fokker-Planck equation we 
investigate the transport of an overdamped colloidal particle in a static, asymmetric periodic potential supplemented by a 
time-dependent, delayed feedback force, $F_{\mathrm{fc}}$.
For a given time $t$, $F_\mathrm{fc}$ depends on the status of the system at a previous time $t-\tau_\mathrm{D}$, with $\tau_\mathrm{D}$ being a delay time, specifically on the delayed mean particle displacement (relative to some ``switching position").
For non-zero delay times $F_{\mathrm{fc}}(t)$ develops nearly regular oscillations generating a net current in the system.
Depending on the switching position, this current is nearly as large or even {\em larger}
 than that in a conventional open-loop rocking ratchet. We also investigate thermodynamic properties of the delayed non-equilibrium system and we suggest an underlying Langevin equation which reproduces the Fokker-Planck results.
 \end{abstract}
\pacs{
  05.40.Jc, 
  05.60.-k, 
  02.30.Yy, 
  05.70.Ln, 
}
\maketitle

\section{Introduction 
\label{SEC:INTRO}}
In recent years, feedback control of transport in small nonequilibrium systems such as ``stairclimbing'' colloids \cite{toyabe10} and fluctuating photon states in quantum-optical systems \cite{sayrin11} has become a topic of active research \cite{sagawa12}.  By definition, ``feedback'' (or closed-loop) control means that the system evolves dynamically under a protocol which depends on an internal variable containing information about the system \cite{bechhoefer05}. It has been shown
that feedback control can lead to pronounced changes of the dynamics compared to purely external (``open-loop'') control and can, in some cases, strongly improve transport properties
such as effective currents. A prime example from the classical side are ratchet systems (or Brownian motors) \cite{haenggi09,Reimann}, specifically the so-called flashing ratchets that
operate by switching on and off a spatially periodic asymmetric
potential: here it has been shown, both theoretically \cite{cao04,feito07,Craig08} and experimentally \cite{lopez08}, that the fluctuation-induced directed transport can be strongly enhanced by
switching not under an externally defined protocol, but ``on demand". Besides the dynamics itself, another topic of intense research is the impact of feedback control on non-equilibrium {\em thermodynamics} \cite{sagawa12,abreu12,Seifert}, concerning particularly the entropy production and fluctuation theorems in both, classical \cite{abreu12,mandal12} and quantum systems \cite{esposito12,strasberg13}. For example, for a classical ratchet model it has been shown that the energy input (work) is smaller with closed- than with open-loop control \cite{sagawa12}.
Exploring these ideas is fostered by recent advancements of experimental techniques for single-particle manipulation and electronic transport, which is of major relevance in various areas such as microfluidics \cite{qian13}, biomedical engineering \cite{fisher05}, and quantum optics \cite{sayrin11}.

Whereas many earlier studies of feedback-controlled systems focused on {\em instantaneous} feedback (i.e., no time lag between measurement and control action) \cite{cao04,brandes10}, there is increasing interest in exploring systems with time delay \cite{feito07,Craig08,Craig,cao12,emary13,munakata14}. The latter typically 
arises from a time lag between the detection of a signal and the control action, an essentially omnipresent situation in experimental setups.
Traditionally, time delay was often considered as a perturbation; for example, in some ratchet systems it reduces the efficiency of transport \cite{feito07}. However, 
time delay can also have significant positive effects. For example, it can stabilize desired stationary states in sheared liquid crystals \cite{strehober13}, it can optimize electron transport
in quantum-dot nanostructures \cite{emary13}, and it can generate new effects such as current reversal \cite{hennig09,lichtner10} and spatiotemporal oscillations in extended systems \cite{lichtner12,gurevich13}. Moreover, 
time delay can have a {\em stabilizing} effect on chaotic orbits, a prime example being Pyragas' control scheme \cite{pyragas92}  of time-delayed feedback control \cite{schoellbuch}.

In this spirit, we discuss in the present paper a classical transport system with feedback, where it is the time delay, which {\em generates} current. Specifically, we consider a so-called rocking ratchet where an overdamped colloidal particle is subject to a combination of a static, asymmetric potential and a time-dependent driving force \cite{Reimann}. This contrasts flashing ratchets, where there is only one type of potential, that is, the asymmetric potential which is switched on 
either periodically (open-loop control) or measurement-dependent (closed-loop control). In a previous study,
Feito {\em et al.} \cite{Feito09} have considered a feedback-controlled flashing ratchet with additional periodic drive. Our feedback system is somewhat closer to the original rocking ratchet model where the total conservative force is the sum of a purely space-dependent and a purely time-dependent force, the latter being typically an oscillation with an externally fixed frequency. 
Contrary to that, the feedback force introduced in Sec.~\ref{model} of this paper depends on the mean particle position, i.e., an internal variable of the system, relative to some reference position in the system. Moreover, we choose
the mean particle position at an {\em earlier} time $t-\tau_{\mathrm{D}}$ as the control target. We show that, due to the time delay $\tau_{\mathrm{D}}$, the feedback force develops an oscillatory behavior which eventually leads 
to a non-zero net current. We also demonstrate that, for appropriate values of the reference position, transport is even {\em improved} as compared to that in a corresponding open-loop device.
Finally, we briefly discuss the entropy production in our system. Indeed, the interplay of (non-equilibrium) thermodynamics \cite{Seifert} and feedback control is a topic attracting
strong interest \cite{sagawa12}, recently also for systems with time delay \cite{munakata14,munakata09,Jiang11}.

Our study is based on a Fokker-Planck equation (FPE) \cite{Risken} where the delayed force enters {\em ad hoc}. 
We note that, in presence of time delay, the connection between the FPE and the underlying Langevin equation is not straightforward (see, e.g., Refs.~\cite{Guillouzic99,Frank03,zeng12}).
Still, since we consider the mean particle position as control target, the results become consistent with those from a corresponding Langevin equation (with delayed force), if the number of noise realizations goes to infinity. In the Appendix of the paper we
demonstrate this consistency numerically and present a justification.

\section{Definition of the model}
\label{model}
We consider the motion of an overdamped colloidal particle at temperature ${\cal T}$ in a one-dimensional, periodic potential $V(z)$, where $z$ is the particle's position. In addition to thermal fluctuations, the particle experiences a time-dependent force $F(t)$. The dynamics is investigated via the FPE \cite{Risken} for the probability density 
\begin{eqnarray}
	\label{FPE}
	\partial_t \rho(z,t) & = &
			\partial_z\left[\gamma^{-1}(V'(z)-F(t))\rho(z,t)+D_0\partial_z\rho(z,t)\right]
		\nonumber
	\\
	& = & -\partial_z G(z,t),
\end{eqnarray}
where $D_0$ is the short-time diffusion coefficient, satisfying the fluctuation-dissipation theorem \cite{Risken}
$D_0=k_{\mathrm{B}}{\cal T}/\gamma$ (with $k_{\mathrm{B}}$ and $\gamma$ being the Boltzmann and the friction constant, respectively), and $G(z,t)$ is the probability current.

%
%

\begin{figure}[htb]
\includegraphics[width=\linewidth]{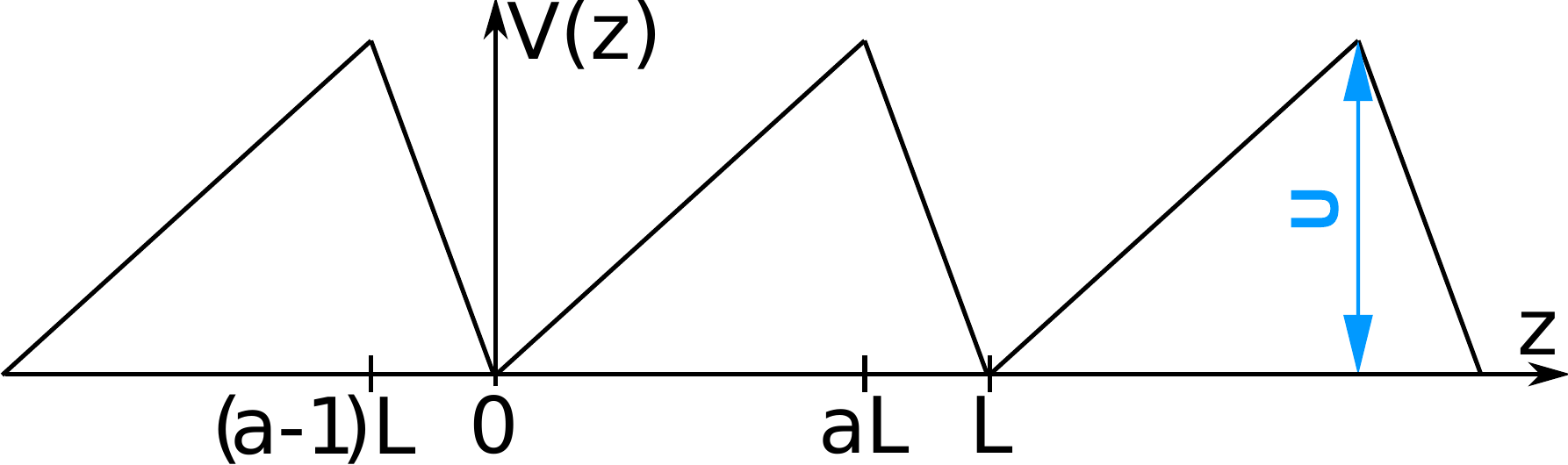}
\caption{(Color online) Sketch of the static ``sawtooth" potential defined in Eq.~(\ref{Ratchet}). The central interval is defined by $(a-1)L\leq z< aL$.}
\label{sketch_V}
\end{figure}

In the present paper we model $V(z)$ by a periodic, piecewise linear, ``sawtooth" potential \cite{lopez08,Kamagawa98,Marquet02} defined by $V(z+L)=V(z)$ and
%
%
\begin{align}
	V(z)=
	\begin{cases}
		Uz/(aL), & 0<z \leq aL,\\
		Uz/((a-1)L) , & (a-1)L < z \leq 0
		\,,
	\end{cases}
	\label{Ratchet}
\end{align}
where $U$ is the potential height, $L$ is the period, and $a$ $\,\in[0,1]$ is the asymmetry parameter. Here we choose $L=8\sigma$, where $\sigma$ is the
diameter of the colloid, and $a=0.8$. The potential minimum is at $z=z_{\mathrm{min}}=0$. An illustration of $V(z)$ is given in
Fig.~\ref{sketch_V}.

In the absence of any further force beyond that arising from $V(z)$, the system approaches for $t\to\infty$ an equilibrium state and thus there is no transport (i.e., no net particle current).
It is well established, however, that by supplementing $V(z)$ by a time-dependent oscillatory force (yielding a ``rocking ratchet"), 
the system is permanently out of equilibrium 
and macroscopic transport can be achieved \cite{Reimann,Magnasco93,Bartussek}. This occurs even when the time-average of the oscillatory force is zero, a characteristic feature
of a true thermal ratchet. 

Here we propose an alternative driving force, where the time dependency arises only through the {\it internal state} of the system. Thus, the force applies feedback control 
onto the system. As a ``control target" we consider the mean particle position within the central interval $S=[(a-1)L,aL[$ 
\begin{gather}
\label{z_average}
		\bar{z}(t)=\int_S\mathrm{d}z\,\rho(z,t)\,z
		\,,
\end{gather}
where $\rho(z,t)$ is the probability density calculated with periodic boundary conditions, that is,
$\rho(z+L,t)=\rho(z,t)$.

Our reasoning behind choosing the {\it mean} rather than the true position as 
control target is twofold: First, within the FPE treatment we have no access to the particle's position
for a given realization of noise, because the latter has already been averaged out. 
This is in contrast to previous studies using Langevin equations \cite{Craig,lopez08,Feito09} where the dynamical variable is the particle position itself. Second, the mean position is an experimentally accessible quantity, which can be monitored, e.g., by video microscopy \cite{Craig}.

Our ansatz for the force reads
\begin{eqnarray} 
\label{f_FC}
F_{\mathrm{fc}}(t) = -F\cdot \mathrm{sign} ( \bar{z}(t-\tau_{\mathrm{D}}) - z_0 )
\,,
\end{eqnarray}
where $F$ is the amplitude (chosen to be positive), $z_0$ is a fixed position within the range $[0,aL]$ (where $V$ increases with $z$), and the sign function
is defined by $\mathrm{sign}(x)=+1$ ($-1$) for $x>0$ ($x<0$). 
From Eq.~(\ref{f_FC}) on sees that the feedback force changes its sign whenever the delayed mean particle position $\bar{z}(t-\tau_{\mathrm{D}})$
becomes smaller or larger than $z_0$; we therefore call $z_0$ the ``switching" position.

Our ansatz is partially motivated by an earlier (Langevin equation based) study of Craig {\em et al.} \cite{Craig} on feedback control
of a flashing ratchet via the so-called ``maximum-displacement strategy". In that study, the fixed position $z_0$ was identified with 
the mean particle position of the {\em uncontrolled system} [i.e., $F_{\mathrm{fc}}(t)=0]$ at $t\to\infty$, that is, the equilibrium position 
$\bar{z}_\mathrm{eq}\!\!=\!\!\int\! \mathrm{d}z z \rho_\mathrm{eq}(z)$, where $ \rho_{\mathrm{eq}}(z)\propto\exp\left[-V(z)/k_{\mathrm{B}}{\cal T}\right]$.
Here we rather regard $z_0$ as a free parameter.
 
Another main feature of our driving mechanism (not considered in Ref.~\cite{Craig}) is the presence of a time delay, $\tau_{\mathrm{D}}$. As discussed in several studies 
(see, e.g., Refs.~\cite{feito07,Craig08,Craig,cao12,emary13,munakata14}), time delay is a rather natural phenomenon which may arise, e.g., through the finite time required for measuring or processing information from a measurement. In the present case, as we will demonstrate below, the time delay is indeed {\it crucial} for generating particle transport.

\section{Transport mechanism}
\label{mechanism}
\begin{figure}
\centering
\includegraphics[width=\linewidth]{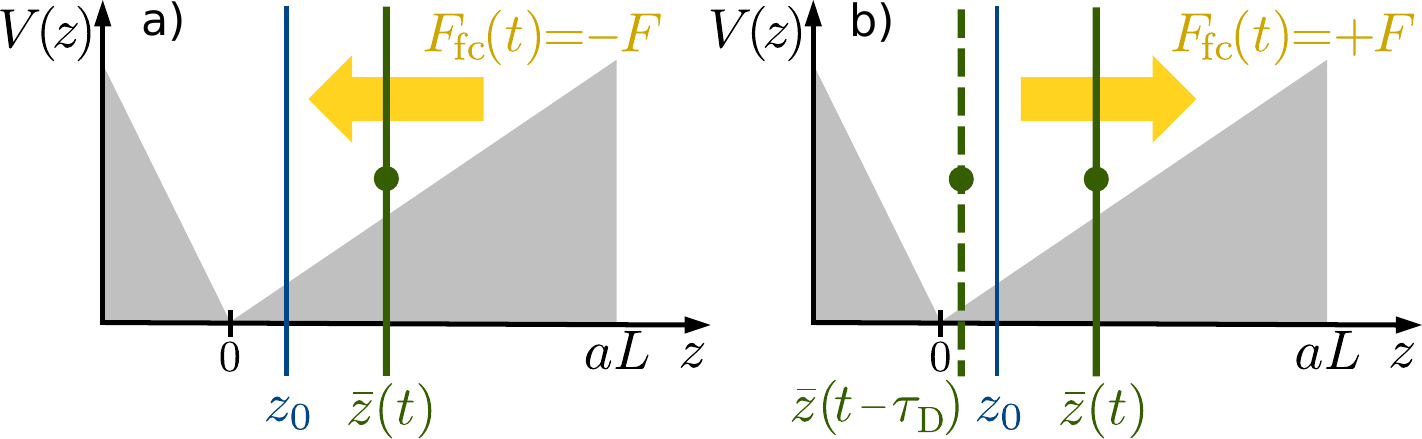}
\caption{(Color online) Sketch of the static potential and the direction of the force $F_{\mathrm{fc}}(t)$ a) in the absence of time delay ($\tau_{\mathrm{D}}=0$), b)
with time delay. The vertical lines indicate the switching position (set to $z_0=0.32\sigma$), as well  
as $\bar{z}(t)$ [and $\bar{z}(t-\tau_{\mathrm{D}})$ in b)].}
\label{Schematic}
\end{figure}

To better understand the impact of the force~(\ref{f_FC}), let us briefly consider the case $\tau_{\mathrm{D}}=0$. For simplicity, we set 
$z_0=\bar{z}_{\mathrm{eq}}\approx 0.32\sigma$.
In Fig.~\ref{Schematic}a) we illustrate a situation, where the mean particle position at time $t$ is on the right hand side of $z_0$.
In this case $F_{\mathrm{fc}}=-F$, meaning that the force tends to push the particle towards $z_0$. In analogous manner we find that $F_{\mathrm{fc}}=+F$ if the particle is left from $z_0$. As time is progressing the mean particle position
thus becomes ``trapped" at $z_0$. Clearly, this excludes any net transport.

However, transport can be generated in the presence of a non-zero time delay, $\tau_{\mathrm{D}}>0$. 
Figure~\ref{Schematic}b) shows as an example a situation where the mean particle position at time $t$ is at the right side of $z_0$, while
it has been on the left side at time $t-\tau_{\mathrm{D}}$. 
In this situation the force $F_{\mathrm{fc}}(t)$ points {\it away} from $z_0$ (i.e., $F_{\mathrm{fc}}>0$), contrary to the case 
$\tau_{\mathrm{D}}=0$ considered in Fig.~\ref{Schematic}a). Thus, the particle experiences a driving force towards the next potential valley, 
which changes only when the delayed position becomes larger than $z_0$. 
The force then points to the left until the delayed position crosses $z_0$ again. This oscillation of the force
(see also Sec.~\ref{target}), together with the asymmetry of $V(z)$, creates a ratchet effect. 

We note that the feedback-controlled ratchet introduced here strongly differs from previous models of such systems. In particular, Feito {\em et al.} \cite{Feito09} have considered
a rocking ratchet composed of a static potential similar to ours plus an oscillatory drive. Feedback-control (based on the average particle force) 
is then introduced as a prefactor in front of the static potential; i.e., the latter is switched on only if the force satisfies certain requirements. 
%
%
In the present model, the control force acts in addition to the static potential, and there is no additional oscillating force.

Another, somewhat subtle aspect of the present model is that we introduce feedback on the level of the Fokker-Planck equation describing the evolution of the probability density. 
This is different from earlier studies
based on the Langevin equation (see, e.g., \cite{Craig,lopez08,Feito09}), where the feedback is applied directly to
the position of one particle, $\chi_i(t)$, or to the average of $N$ particle positions $N^{-1}\sum_{i=1}^{N} \chi_i(t)$. Introducing feedback control in such systems
implies to introduce effective {\em interactions} between the particles. As a consequence, 
the transport properties in these particle-based models depend explicitly on the number of particles, $N$. 
Typically it turns out that the current becomes small or even vanishes when the particle number increases, the reason being that fluctuations (which are essential for the ratchet effect) disappear
\cite{cao04}.
From the perspective of these Langevin-based models,
the present model corresponds to the ``mean-field'' limit $N\to\infty$.
This connection to a Langevin model is further discussed in the Appendix. Given that we are in the ``mean-field'' limit, it is even more interesting that we do observe
a non-vanishing current which can be even larger than in an open-loop system \cite{cao04}. This is because our model involves a time delay.

\section{Numerical results}
\label{results}
\subsection{Dynamics of the control target}
\label{target}
In this section we present numerical results for the feedback-controlled transport based on numerical solution of the FPE Eq.~(\ref{FPE}). 
The height of the static potential is set to $U=15k_{\mathrm{B}}{\cal T} $. In fact, similar values are found in experiments of colloids in structured light fields
\cite{lopez08,dalle11,evstigneev08}.
Time is measured in units of the Brownian timescale, $\tau=\sigma^2/D_0$,
which is of the order of $10^0s$ to $10^2s$ for typical colloids \cite{lopez08,dalle11,lee05,evstigneev08,tierno10}.
In all calculations, the initial condition for the probability density is a $\delta$-function localized at the minimum of $V(z)$, $z_{\mathrm{min}}=0$.
Further, to initialize the control force, we set $\bar{z} (t)=z_{\mathrm{min}}=0$ for $t\in\lbrack -\tau, 0 \rbrack$. 
The data presented in the following correspond to time ranges after an initial (yet very short) ``equilibration" period, after which
the dynamic quantities considered display a regular dynamical behavior.

\begin{figure}
\centering
\includegraphics[width=\linewidth]{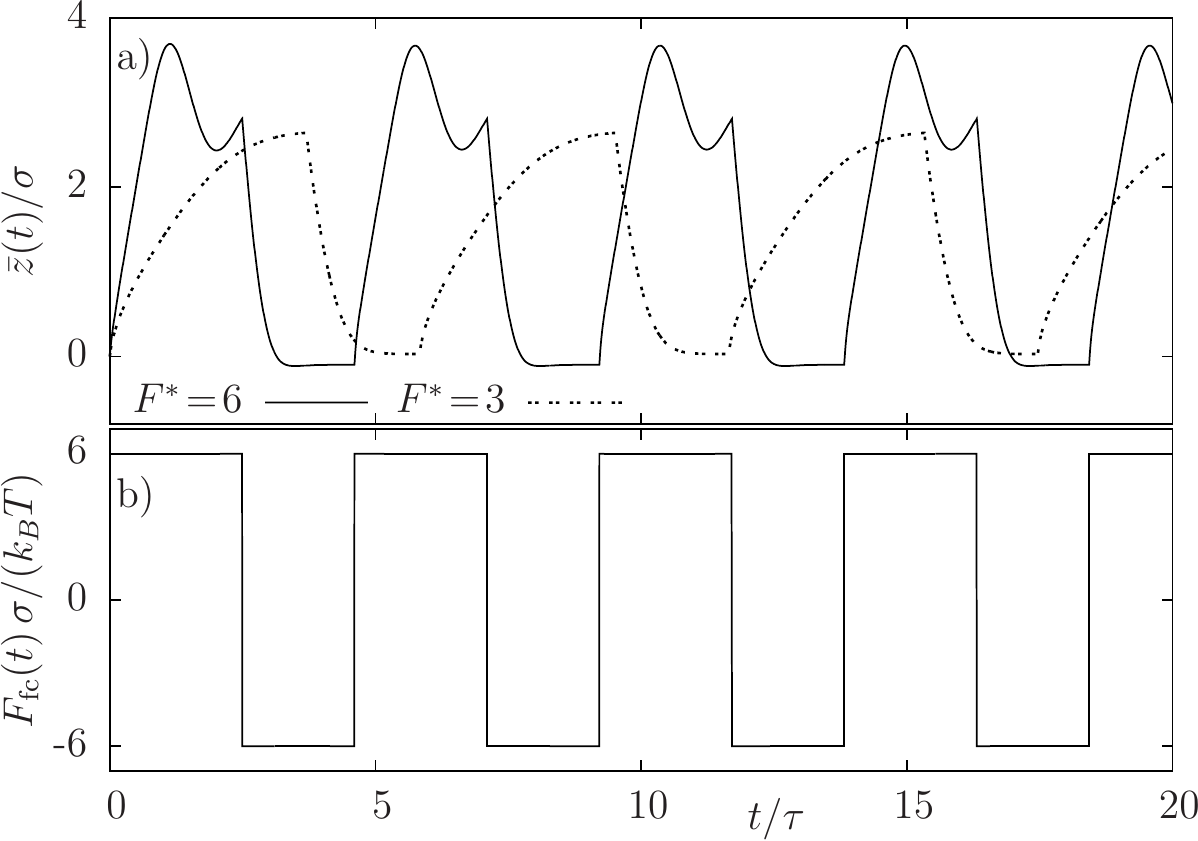}
\caption{(Color online) a) Mean particle position and b) control force as functions of time for $\tau_{\mathrm{D}}=2\tau$, $z_0=2\sigma$ and a) $F^{*}\in\{3,6\}$ and b) $F^*=6$.}
\label{mz}
\end{figure}

We start by considering the time evolution of the mean particle position, $\bar{z}(t)$, which determines the control force. 
Exemplary data 
for two amplitudes $F^{*}=F\sigma/k_{\mathrm{B}}{\cal T}$ are shown in Fig.~\ref{mz}a), where the parameter $z_0$ has been set to $2\sigma$, and $\tau_{\mathrm{D}}=2\tau$. 
It is seen that
$\bar{z} (t)$ displays regular oscillations between values above and below $z_0$ for both force amplitudes considered. The period of these oscillations, $\bar{T}$, 
is roughly twice the delay
time. 
However, the precise value of the period as well as the shape of the oscillations depend on the values of $F^{*}$ and $z_0$ (see also Sec.~\ref{current}). 

Due to the oscillatory behavior of $\bar{z}(t)$, the delayed position $\bar{z}(t-\tau_{\mathrm{D}})$ oscillates around $z_0$ as well. 
It follows from our definition of the feedback force [see Eq.~(\ref{f_FC})], that the latter switches periodically between $+F$ and $-F$ with the same period
as that observed in $\bar{z} (t)$. This is clearly seen in Fig.~\ref{mz}b) where we plotted $F_{\mathrm{fc}}(t)$ for the case $F^{*}=6$.

\begin{figure}
\centering
\includegraphics[width=\linewidth]{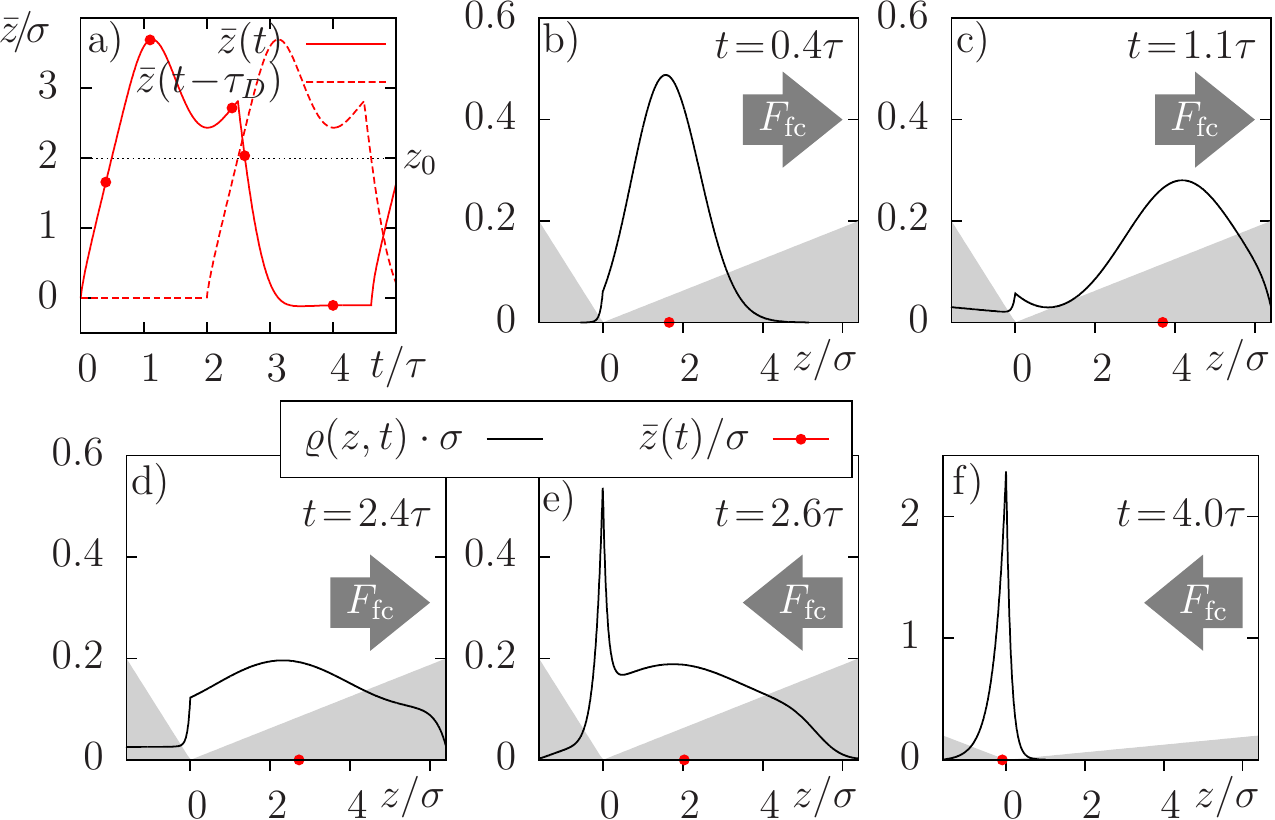}
\caption{(Color online) a) One cycle of the function $\bar{z}(t)$ at $\tau_{\mathrm{D}}=2\tau$ and $F^{*}=6$, with the filled circles indicating specific times. The switching position is set to
$z_0=2\sigma$. Also shown is the corresponding function
$\bar{z}(t-\tau_{\mathrm{D}})$.
b)-f) Density distribution as function of space at the times indicated in a). The thick arrows show the direction of the control force. The filled (red) circles indicate the values of $\bar{z}(t)$.}
\label{cycle}
\end{figure}

A closer view on the dynamic behavior within one cycle is given in Fig.~\ref{cycle}, where we focus on the case $\tau_{\mathrm{D}}=2\tau$ and $z_0=2\sigma$.
Figure~\ref{cycle}a) depicts one cycle of the function $\bar{z}(t)$ together with its time-delayed counterpart, $\bar{z}(t-\tau_{\mathrm{D}})$.
The remaining parts of Fig.~\ref{cycle} then plot the probability density $\rho$ as function of $z$ for specific times indicated
by filled circles in Fig.~\ref{cycle}a). 

The mean particle position starts from $\bar{z}=0$ (i.e., localization in the potential minimum) at $t_0=0$.
The amplitude $F^*$ is chosen large enough ($F^{*}=6$) so that the total systematic force $-V'(z)+F_\mathrm{fc}$ at $t=0$ is positive for every $z$. Hence,
at $t=0.4\tau$ [part b)], the density distribution has broadened by diffusion and $\bar{z}(t)$ has moved to the right.
We also see from Fig.~\ref{cycle}b) that the probability density is still very small at the boundary. This changes at $t=1.1\tau$ when probability ``flows'' over the boundary, indicating transport [see part c)]. The function $\bar{z}(t)$ is now in its maximum. At $t=1.1\tau$ the mean particle position has already crossed $z_0$; however, the time delayed mean particle position is still below $z_0$, and thus, $F_{\mathrm{fc}}(t)>0$. Note that due to the periodicity of the system an inward probability flow occurs at the lower boundary. With progressing time this eventually leads to a shift of the mean particle position towards smaller values, as seen in Fig.~\ref{cycle}d) for the case $t=2.4\tau$.
When $\bar{z}(t-\tau_\mathrm{D})$ crosses $z_0$ the feedback control force is reversed. The total systematic force is now positive for $z<0$ and negative for $z>0$. The confining effect of this force to the particle can be seen in Fig.~\ref{cycle}e) where a peak in the probability density evolves. 
As a consequence, the mean particle position moves towards values around the potential minimum.
When the same happens to the delayed position $\bar{z}(t-\tau_\mathrm{D})$, the cycle starts again.

We remark that in order to see {\em persistent} oscillations of the control target and thus, the control force, it is crucial
that the function contained in $F_\mathrm{fc}$ is very sensitive to even tiny 
differences between  $\bar{z}(t-\tau_{\mathrm{D}})$ and $z_0$. Indeed, besides the sign-function we have also tested continuous functions such as $\sin(x)$ [or $\cos(x)$], which 
tend to zero when $\bar{z}(t-\tau_{\mathrm{D}})-z_0\rightarrow 0$ [or $\pi/2$]. In these cases, the oscillations
just dampen out and thus, there is no ratchet effect.
\subsection{Effective current}
\label{current}
So far we have focused on the mean particle position $\bar{z}(t)$ within one interval [see Eq.~(\ref{z_average})], i.e., the quantity determining our feedback force. However, to visualize the particle transport, it is more convenient to consider the distance $\tilde{z}(t)$ the particle has actually traveled at time $t$ (in the ensemble average) relative to its value at $t=t_0$. 
Contrary to $\bar{z}(t)$, the travelled distance $\tilde{z}(t)$ takes into account that the particle actually moves from one potential valley to the next.

To this end we first introduce the particle current
\begin{align}
	\label{j_current}	 
		j(t)=\int_{S}\mathrm{d}z\,G(z,t)
		\,,
\end{align}
with the probability current $G(z,t)$ calculated from the FPE~(\ref{FPE}) with periodic boundary conditions. As shown 
in Ref.~\cite{Reimann}, this current can also be expressed as 
\begin{align}
	\label{j_intermediate}
		j(t)
	&
		=\frac{\mathrm{d}}{\mathrm{d}t}\left[\int_{z_\mathrm{ref}}^{z_\mathrm{ref}+L}\!\!\!\!\mathrm{d}z\,z\,\rho(z,t)\right]+L\,G(z_\mathrm{ref},t)
	\nonumber\\
	&
		=\frac{\mathrm{d}}{\mathrm{d}t}\bar{z}(t)+L\,G(z_\mathrm{ref},t)
		\,,
\end{align}
%
where $z_{\mathrm{ref}}$ is an arbitrary reference position within the central interval. Here we choose $z_{\mathrm{ref}}$ equal to $(a-1)L$, that is, the lower boundary of
the central interval.
Equation~(\ref{j_intermediate}) expresses the
fact that the particle current is composed of the motion of the ``center of mass" plus $L$ times
the probability current (evaluated for the periodic system) at the reference point.
We now define $\tilde{z}(t)$ as the time integral of $j(t)$, yielding
\begin{gather}
	\label{distance}
		\tilde{z}(t)=\int_{t_0}^t \mathrm{d}t'\,j(t')=\bar{z}(t)+L\int_{t_0}^t \mathrm{d}t'\,G(z_{\mathrm{ref}},t')
		\,,
\end{gather}
where we have used that $\tilde{z}(t_0)=\bar{z}(t_0)$. 

\begin{figure}
\centering
\includegraphics[width=\linewidth]{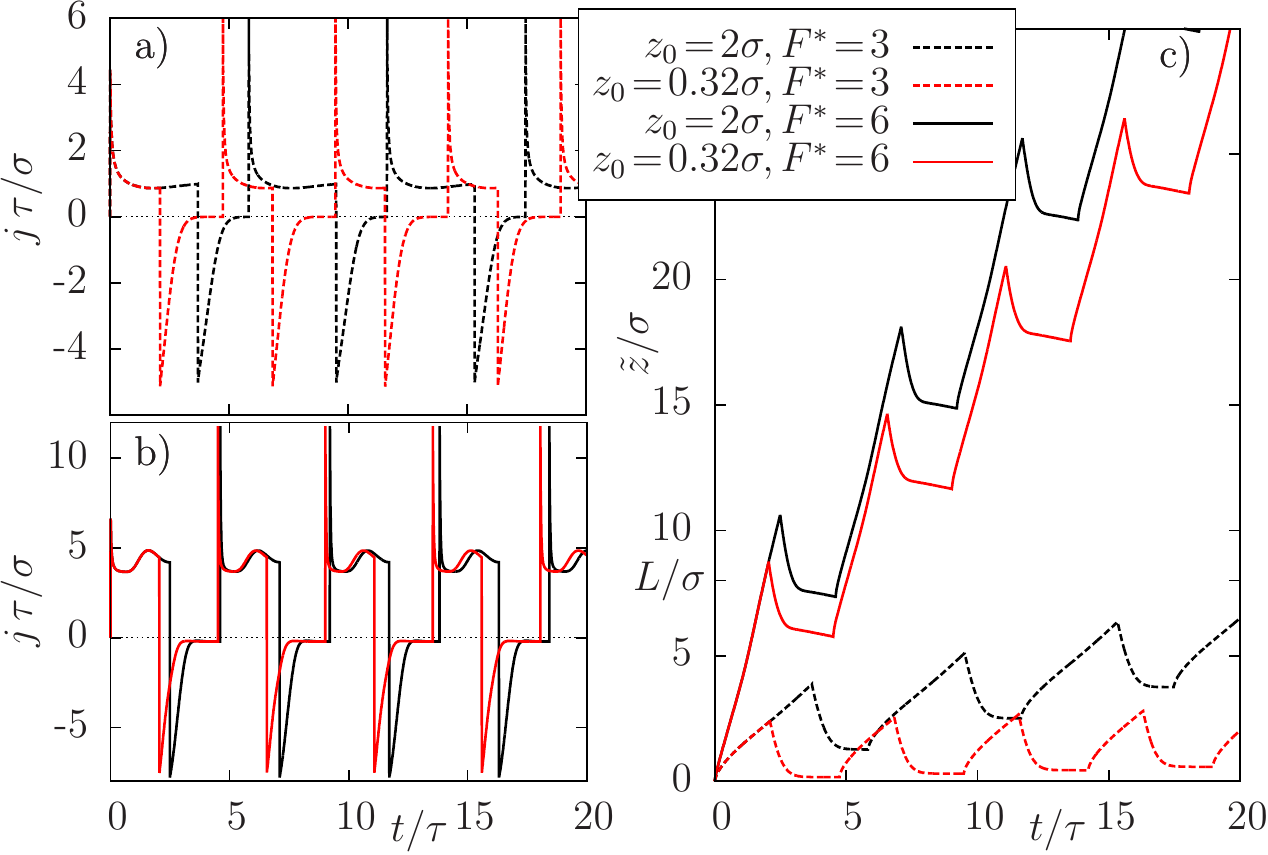}
\caption{(Color online) a), b) Space-averaged current density $j$ as function of time for a) $F^{*}=6$ and b) $F^{*}=3$ (and different switching positions). c) Travelled distance $\tilde{z}$ as function of time. In all parts the delay time is set to $\tau_{\mathrm{D}}=2\tau$.}
\label{actual_position}
\end{figure}

Numerical results for $\tilde{z}(t)$ and $j(t)$ are plotted in Fig.~\ref{actual_position} for different values of the control force parameters $F^{*}$ and $z_0$. The delay time is kept fixed. 
In all cases considered,
$\tilde{z}(t)$ displays a regular ``back-forth" rocking motion, but with a net drift to the right. 
The latter indicates that there is indeed particle transport. Also shown in Fig.~\ref{actual_position}
is the space-averaged current defined in Eq.~(\ref{j_current}).
It is seen that $j(t)$ reflects the rocking-like behavior of $\tilde{z}(t)$ by oscillations around zero. 
The fact that the positive values in $j(t)$ dominate signals the presence of net transport.

\begin{figure}
\centering
\includegraphics[width=\linewidth]{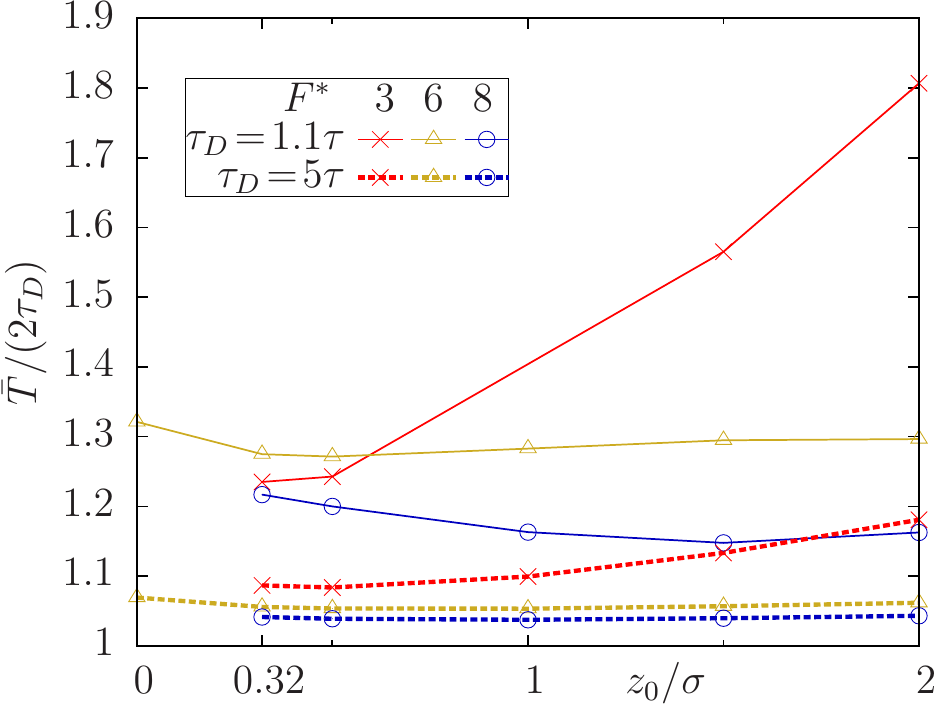}
\caption{(Color online) Mean period of the oscillations of the control target in dependence on $z_0$ and $F^{*}$ (for $\tau_{\mathrm{D}}\in\{1.1\tau,5\tau\}$).}
\label{period}
\end{figure}

Not surprisingly, both the current and the strength of the drift visible in $\tilde{z}(t)$ depend on the amplitude of the control force, as one can clearly see by comparing
the dashed and solid curves in Fig.~\ref{actual_position}. However, we also observe a significant influence of the position $z_0$:
The larger $z_0$, the longer are the times in which the travelled distance increases in each cycle and in which the current is positive.
We understand this behavior such that, the larger $z_0$, the longer the time in which the mean particle position in the central interval $S$ is below $z_0$ (yielding $F_{\mathrm{fc}}>0$).
The period $\bar{T}$ of the oscillations of $\tilde{z}(t)$ and $j(t)$ 
slightly increases with $z_0$ as well. An overview of the dependence of $\bar{T}$ on $z_0$ and $F^{*}$ is given in Fig.~\ref{period}. 
In all cases, $\bar{T}$ is {\em roughly} given by twice the
delay time, however, its actual value depends on the precise choice of the control force parameters. 

\begin{figure}
\centering
\includegraphics[width=\linewidth]{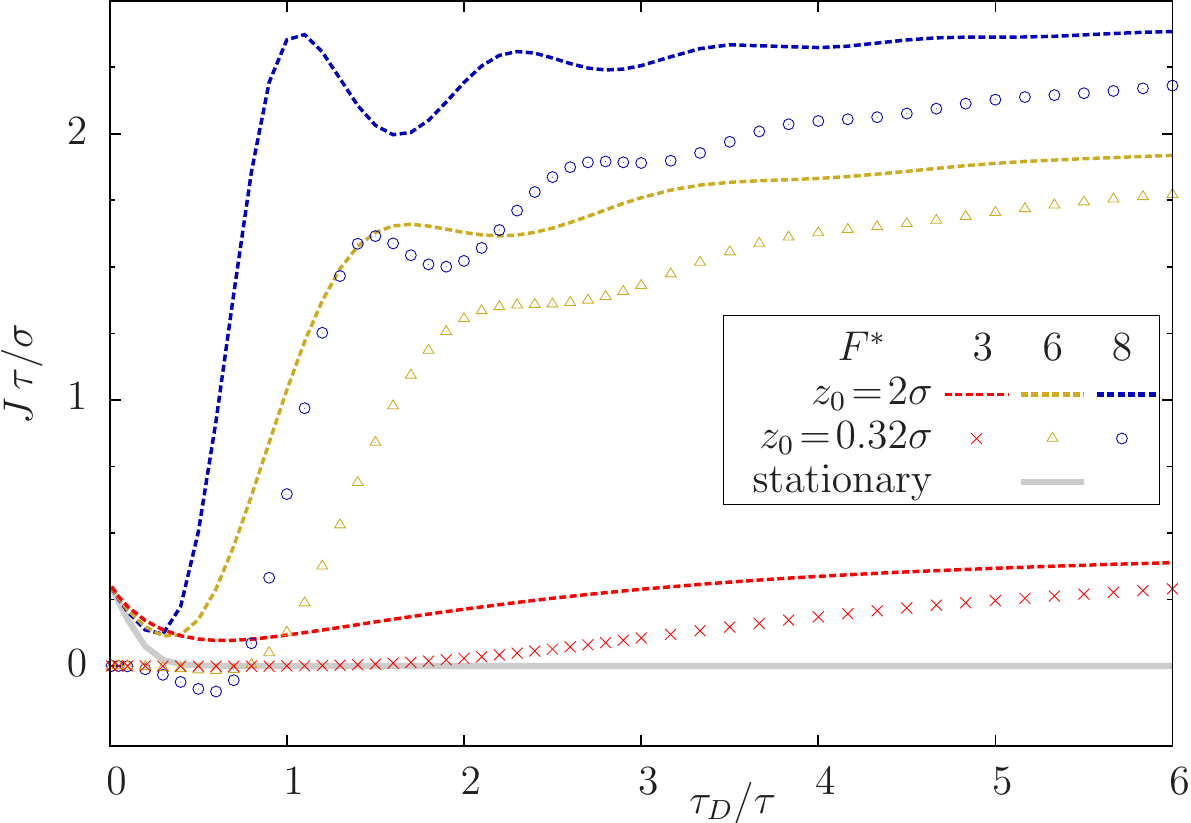}
\caption{(Color online) Net particle current $J$ as function of the delay time for several values of $F^{*}$ and $z_0$. The curve termed ``stationary'' shows the time-constant current $j$ 
in a system with constant force $\bar{F}_\mathrm{fc}(\tau_\mathrm{D},z_0\!=\!2\sigma,F^*\!=\!6)$.}
\label{J_delay}
\end{figure}

Having understood the time-dependence of the control target and the current density we now turn to the overall (time-averaged) transport. The latter is
measured by the net particle current defined as
\begin{gather}
\label{J}
	J=\bar{T}^{-1}\int_{t_1}^{t_1+\bar{T}}\mathrm{d}t'\,j(t')
	\,,
\end{gather}
where $j(t)$ is defined in Eq.~(\ref{j_current}), and $t_1$ is an arbitrary time after the ``equilibration" period. 
Numerical results for $J$ in dependence of the delay time $\tau_{\mathrm{D}}$ and force amplitude $F^{*}$
are plotted in Fig.~\ref{J_delay}, where we consider two switching positions.

We first discuss the behavior at finite delay times in the range $\tau_{\mathrm{D}}\gtrsim 5\tau$. In this range
the current generally increases with $\tau_{\mathrm{D}}$, with the increase 
being the more pronounced the larger the force amplitude and the switching position is. This is consistent with our earlier findings regarding
the particle's travelled distance and the time-dependent current [see Fig.~\ref{actual_position}]. We also see from Fig.~\ref{J_delay} that
all curves saturate in the limit $\tau_{\mathrm{D}}\to\infty$ at some finite value of $J$ which solely depends on $F^{*}$.

At small delay times ($\tau_{\mathrm{D}}\lesssim 5\tau$) the behavior of the function $J(\tau_{\mathrm{D}})$ strongly depends on both, $F^{*}$ and $z_0$.
For $z_0=0.32\sigma$, the net current vanishes at $\tau_{\mathrm{D}}\to 0$ regardless of the strength of the drive, consistent with our previous considerations 
that the ratchet effect in our model is essentially driven by the time delay. Upon increasing $\tau_{\mathrm{D}}$ the current then deviates from zero. Interestingly, for large
force amplitudes ($F^{*}=6$, $8$), $J$ may even become {\em negative} before finally increasing towards positive values. Note that
negative values imply transport {\em opposite} to the direction supported by the asymmetric potential.

\begin{figure}
\centering
\includegraphics[width=\linewidth]{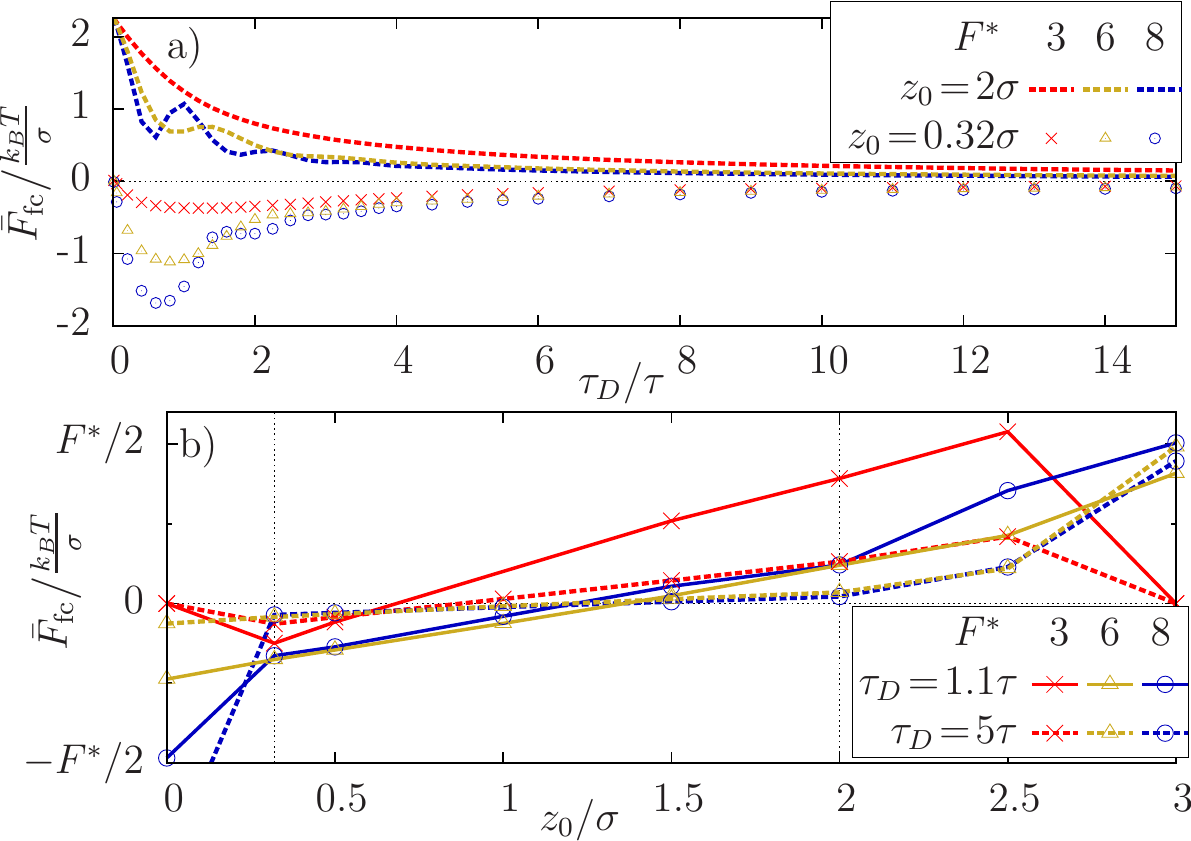}
\caption{(Color online) Time-averaged control force $\bar{F}_{\mathrm{fc}}(t)$ as function of a) the delay time (with $F^{*}=6$ and two values of $z_0$), b) the switching position $z_0$.
In b), the dotted lines indicate the boundaries of the range of switching positions considered in this paper.}
\label{f}
\end{figure}

Considering now the larger switching position $z_0=2\sigma$, we observe again a strong decrease of the current when we decrease the delay times from large values.
However, contrary to the situation at $z_0=0.32\sigma$, $J(\tau_{\mathrm{D}})$ stays {\em finite} in the limit $\tau_{\mathrm{D}}\to 0$.
We can understand this behavior, as well as the negative currents arising at $z_0=0.32\sigma$ and $F^{*}=8$,
by considering the time average of the control force, 
\begin{gather}
		\bar{F}_{\mathrm{fc}}=\bar{T}^{-1}\int_{t_1}^{t_1+\bar{T}}\mathrm{d}t'\,F_{\mathrm{fc}}(t')
		\,.
\end{gather}
Figure~\ref{f}a) plots the averaged control force as function of $\tau_{\mathrm{D}}$. 
Considering first the case $z_0=0.32\sigma$, we see that $\bar{F}_{\mathrm{fc}}(\tau_{\mathrm{D}})$ approaches zero in the limit of vanishing delay time.
In other words, there is no {\it average} drive, which justifies to consider the present transport 
mechanism as a true (delay-induced) ratchet effect.  For small $\tau_{\mathrm{D}}$, however, there is a minimum in the function, which becomes the more pronounced
the larger $F$ is. The negative values of $\bar{F}_{\mathrm{fc}}(\tau_{\mathrm{D}})$ are responsible for the 
negative net current $J$ arising in the same range of delay times (see Fig.~\ref{J_delay}). 
Therefore, the appearance of negative $J$ here has a different origin than in the open-loop controlled case \cite{Bartussek}.

At $z_0=2\sigma$ the average force is non-zero and positive throughout the entire range of delay times, becoming largest in the limit $\tau_{\mathrm{D}}\to 0$. We note, however, that at any finite delay time the absolute value of  $\bar{F}_{\mathrm{fc}}$ is quite small. To check the influence of this remaining force we have calculated the current $J$ for a system
under the time-constant force $\bar{F}_{\mathrm{fc}}$, taking the case
$\bar{F}_{\mathrm{fc}}^{*}=6$ as an example. It turns out that this current is indeed negligible except at $\tau_{\mathrm{D}}\to 0$ (indicated by the curve termed ``stationary'' in Fig.~\ref{J_delay}). Thus, we can conclude that even with this
larger switching position the ratchet effect is essentially delay-induced. A more systematic view of the dependence of $\bar{F}_{\mathrm{fc}}$ on $z_0$ is given
in Fig.~\ref{f}b), where we focus on specific, finite values of $\tau_{\mathrm{D}}$.
It is seen that, outside the range $0.32\sigma \lesssim z_0 \lesssim 2\sigma$ (see vertical dotted lines in Fig.~{\protect\ref{f}}b)),
 the average force deviates significantly from zero. In these cases, it becomes questionable to which extent the current is really induced by time delay. Therefore
 we have restricted $z_0$ to values inside the interval defined above.

\begin{figure}
\centering
\includegraphics[width=\linewidth]{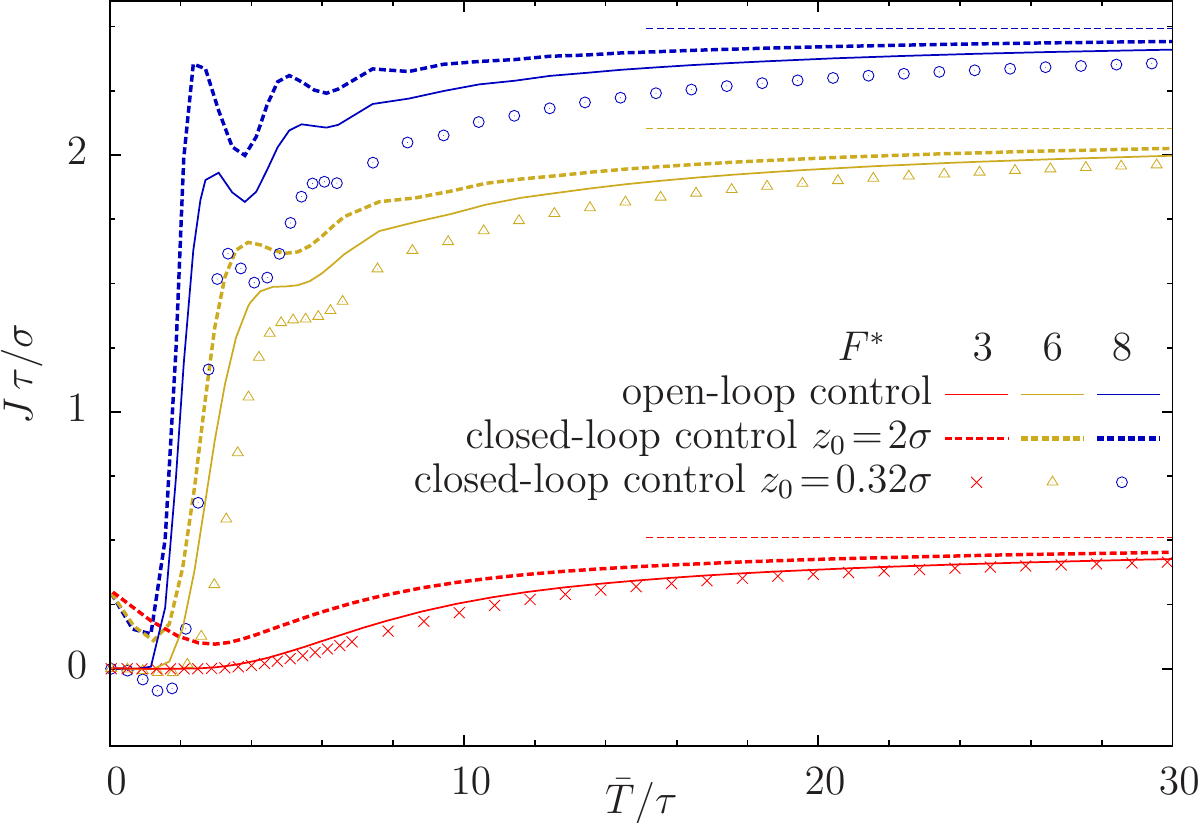}
\caption{(Color online) Net particle current $J$ for the open-loop rocking ratchet and the feedback-controlled ratchet in dependency of the (mean) oscillation period of the driving force. The horizontal lines pertain to the adiabatic limit [see Eq.~(\ref{J_limit})].}
\label{J_period}
\end{figure}

At this point it is worth to compare the current generated by our feedback-controlled ratchet with that of an ordinary, ``open-loop" rocking ratchet. 
To this end we supplement the static periodic potential given in Eq.~(\ref{Ratchet}) by a time-periodic drive characterized by a fixed period $T$ 
with vanishing time average. To be as close as possible 
to our feedback model [see Eq.~(\ref{f_FC})], we choose a {\em rectangular} oscillatory drive
\begin{eqnarray}
\label{os_drive}
F_{\mathrm{osc}}(t)=-F\cdot \mathrm{sign} \left[ \cos{ \left( \frac{2 \pi}{T} t \right) } \right]
\,.
\end{eqnarray}
The resulting net current can be calculated via Eq.~(\ref{J}) after replacing $\bar{T}$ by $T$. 
In Fig.~\ref{J_period} we show numerical results for $J$ as function of the oscillation period, together with the corresponding functions $J(\bar{T})$ for a feedback-controlled
ratchet with two values of $z_0$.

While the general behavior of the current (that is, small values of $J$ for small periods, saturation at large values for large periods) 
is similar for both, open-loop and closed-loop systems, the actual values of $J$ for a given period strongly depend on the type of control. 
This is seen already at very small periods where, e.g., the current of the closed-loop system with $z_0=0.32\sigma$ can become negative,
while that of the open-loop system is still zero. The most interesting differences, however, occur at finite periods which are still below those corresponding to the saturation regime: 
Comparing curves with the same value of $F^{*}$ we find that the net current in the open-loop system
is larger than in the closed-loop system with small switching position ($z_0=0.32\sigma$), but {\em smaller} than in the closed-loop system with $z_0=2\sigma$.
In other words, the net current, which is the measure for transport, can be larger in the feedback-controlled system than that in the open-loop case, provided that the
switching position is sufficiently large. At very large periods, however, the currents corresponding to a given value of $F$ 
approach the same values. The latter correspond to the ``adiabatic limit'' ($T\to\infty$), where the drive changes so slowly so that the system can be assumed to be in a steady state at every time $t$ \cite{Reimann}. This allows to calculate the current analytically, yielding
\begin{gather}
	\label{J_limit}
		J=\frac{D_0L}{T} \int_{0}^{T}\mathrm{d}t\left( 1-e^{-\frac{L F(t)}{k_\mathrm{B}\mathcal{T}}}\right)/\mathcal{N}(t),
\end{gather}
where
\begin{gather}		
		\mathcal{N}(t)=\int_S\mathrm{d}z\,e^{-\frac{V(z)-z F(t)}{k_\mathrm{B}\mathcal{T}}} \int_{z}^{z+L}\!\!\!\mathrm{d}z'\,e^{\frac{V(z')-z'F(t)}{k_\mathrm{B}\mathcal{T}}}
		\,,
	\notag
\end{gather}
and $F(t)=F_{\mathrm{osc}}(t)$ and $F(t)=F_{\mathrm{fc}}(t)$ for the open- and closed-loop case, respectively. 
%

\section{Entropy production and work}
\label{thermo}
In view of our results for the net current in the feedback-controlled ratchet, on the one hand, and the 
open-loop controlled rocking ratchet, on the other hand (see Fig.~\ref{J_period}), it is interesting to further explore the impact of the
control scheme in terms of (nonequilibrium) thermodynamics. In particular, we are interested in the {\em total entropy production}, which measures how far the system is away from equilibrium, and in
the {\em work} that is performed on the particle. We calculate these quantities on the basis of stochastic thermodynamics.  
For systems with {\em instantaneous} feedback control this is a well-established field
\cite{Seifert,sagawa12}. This is generally not the case for systems with time delay, in which the underlying (Langevin or Master) equations of motion become {\em non-Markovian} such that 
concepts of standard stochastic thermodynamics (which assumes
Markovian dynamics) are not immediately applicable \cite{munakata14,Jiang11}.

In the present case the situation is somewhat easier because we are working in a mean-field limit. As discussed in the Appendix, this limit allows us to establish a connection between our FPE 
and an underlying Langevin equation; it also allows us to consider our delayed feedback control force just as a special type of time-dependent force. In the following we stress this argument further and use various FPE-based standard formula for thermodynamic quantities. To test the FPE results we compare with those obtained from trajectory-based expressions via Brownian 
Dynamics (BD) simulations.

We start by considering the {\em averaged} total entropy production, $\dot{S}^{\mathrm{tot}}$. Within stochastic thermodynamics, the total entropy $s^{\mathrm{tot}}(t)$, for a single trajectory $\chi(t)$, 
consists of two contributions \cite{Seifert}, i.e., $s^{\mathrm{tot}}(t)=s(t)+s^{\mathrm{m}}(t)$. Here,
$s(t)=-k_\mathrm{B}\ln \rho(\chi(t),t)\sigma$ is the trajectory-dependent entropy of the ``system'' (i.e., the particle), and $s^{\mathrm{m}}(t)=q[\chi(t)]/\mathcal{T}$ is the medium 
entropy related to the heat $q[\chi(t)]$ dissipated into the medium.
Upon averaging over the ensemble of trajectories \cite{Seifert}, one finds the following compact expression for the time-derivative (production rate) of the total entropy
\begin{equation} 
\label{total_FPE}
\dot{S}^\mathrm{tot}(t)=k_B\int_S\mathrm{d}z\,\frac{G(z,t)^2}{D_0\,\rho(z,t)}\,,
\end{equation}
where $G(z,t)$ is the probability current [see Eq.~(\ref{FPE})]. 

\begin{figure}
\centering
\includegraphics[width=\linewidth]{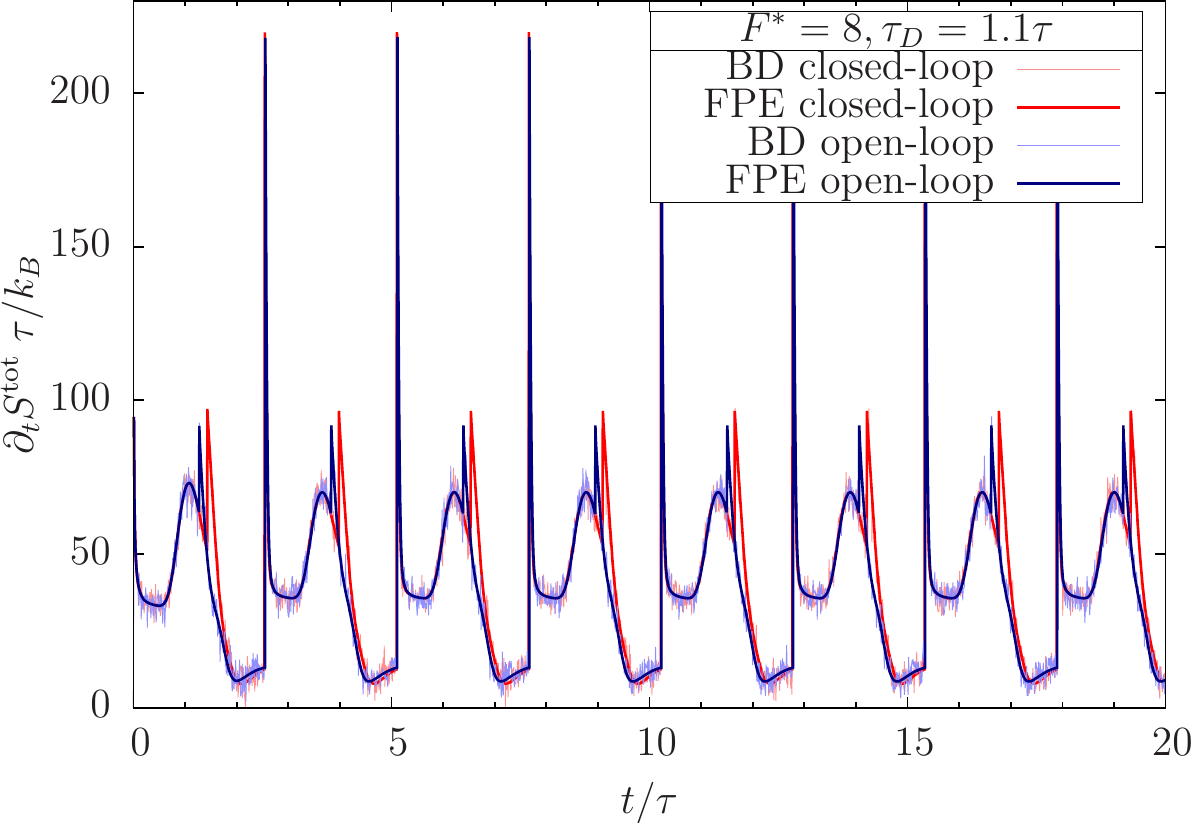}
\caption{(Color online) Total entropy production as function of time for $F^{*}=8$, $\tau_{\mathrm{D}}=1.1\tau$, and $z_0=2\sigma$, and for the corresponding open-loop-controlled ratchet
with $F^{*}=8$ and $T=2.558\tau$. Included are results based on BD simulations of the Langevin equation (see Appendix).}
\label{Entropy_production}
\end{figure}

Numerical results for $\dot{S}^\mathrm{tot}$ are shown in Fig.~\ref{Entropy_production} where we focus on a situation where the net current
in our closed-loop scheme is larger than in the open-loop system. 
Included are results for the corresponding open-loop system (in which $T\colonequals\bar{T}$). For both the closed-loop and the open-loop system, $\dot{S}^\mathrm{tot}(t)$ displays periodic behavior 
with similar features. First, the beginning of a new cycle is indicated by a very large and narrow peak. After the peak $\dot{S}^\mathrm{tot}(t)$ decreases to a small, yet non-zero value and then rises towards a second, broader maximum, followed by a further sharp peak. The latter 
is related to the change of the feedback force from positive to negative values. 
For the open-loop system this happens exactly in the middle of the cycle [see Eq.~\eqref{os_drive}]. In the closed-loop system, the change is somewhat shifted. This deviation is indeed the main difference between the closed-loop- and the open-loop-controlled system. 

We have also calculated the total entropy production by BD simulations based on Eq.~(\ref{Langevin}). On that level, the rate of change of the system entropy is given as 
$\dot{S}(t)=-\mathrm{d}/\mathrm{d}t\langle \ln \rho(\chi(t),t)\sigma\rangle$, with $\langle\ldots\rangle$ being a noise average. In practice, we have evaluated $\dot{S}$ using
the relation $\langle \ln \rho(\chi(t),t)\sigma\rangle=\int_S\mathrm{d}z\,\rho(z,t)\ln\rho(z,t)\sigma$ where the probability density 
is calculated as $\rho(z,t)=\langle \delta\left(z-\chi(t)\right)\rangle$. Further, 
the medium entropy is calculated from \cite{Seifert}
\begin{equation}
\label{entropy_BD}
\dot{S}^{\mathrm{m}}(t)=
\langle
(-V'(\chi(t))+F_\mathrm{fc}^N(t))\dot{\chi}(t)/\mathcal{T}\rangle
\,.
\end{equation}
To evaluate this expression we have used the Stratonovich interpretation. It is seen that the BD data 
(which have been obtained with $N=10^5$) are fully consistent with those from the FPE approach.

\begin{figure}
\centering
\includegraphics[width=\linewidth]{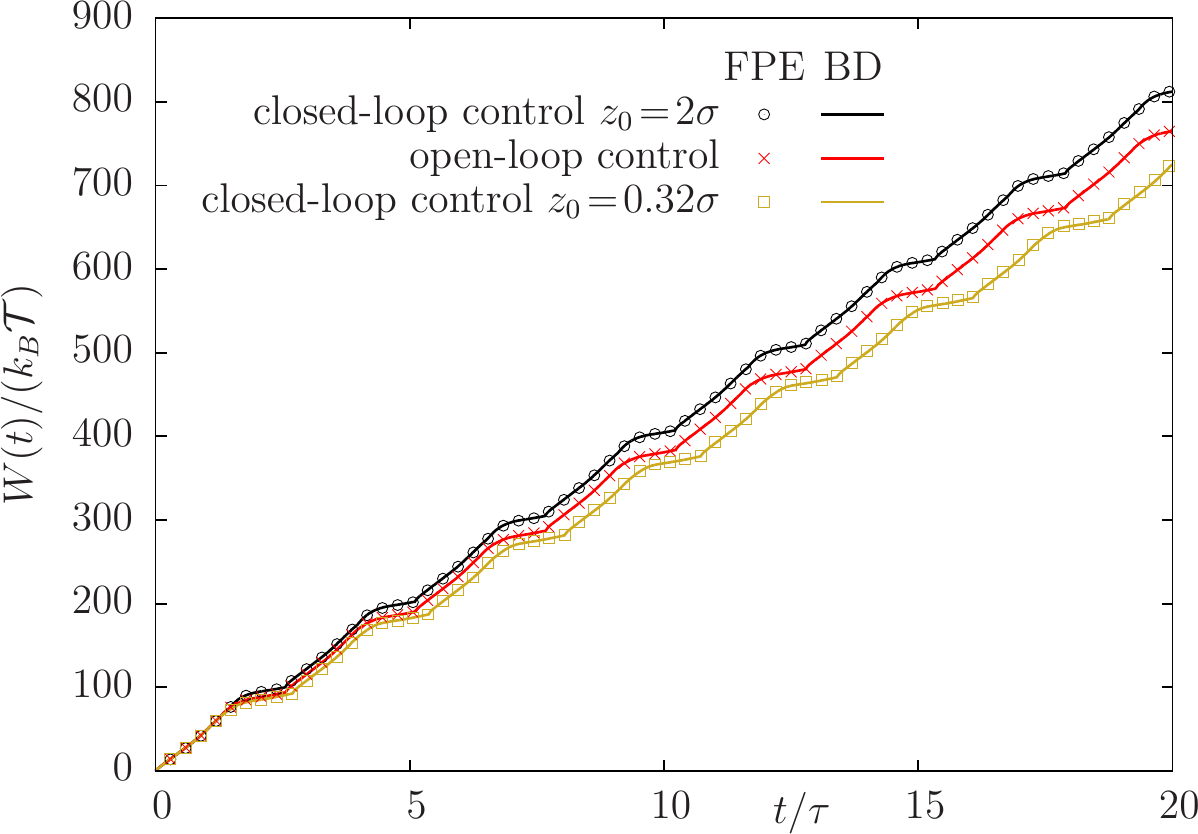}
\caption{(Color online) Work performed on the particle as function of time for $F^{*}=8$, $\tau_{\mathrm{D}}=1.1\tau$, and two switching positions $z_0$. 
Included are results for the corresponding system under open-loop control.}
\label{Work}
\end{figure}

To calculate the work performed on the particle we note that, contrary to the dissipated heat, the work involves only changes of the total systematic force at fixed particle position
\protect\cite{Seifert} which is, in our case, $F_\mathrm{fc}(t)$. On the level of a single trajectory $\chi(t)$ the work therefore reads
\begin{gather}
	\label{work_BD}
		w[\chi(t)]=\int_0^{t}\!\mathrm{d}t'\,F_\mathrm{fc}(t')\,\dot{\chi}(t')
		\,.
\end{gather}
To achieve a description in terms of the FPE we make use of the formula $\langle a(z)\dot{z}\rangle=\int_S\mathrm{d}z\,G(z,t)\,a(z)$ \protect\cite{Seifert} (implicitly 
assuming again that the time delayed feedback control force enters the FPE just like a special time-dependent force). The noise-averaged work is then given by
\begin{equation}
\label{work_FPE}
W(t)=\int_0^t\mathrm{d}t'\,F_\mathrm{fc}(t')\int_S\mathrm{d}z\,G(z,t')\,.
\end{equation}
In Fig.~\ref{Work} we compare the time-dependence of the work for the closed-loop system with two different switching positions with the corresponding open-loop system.
It is seen that the work increases in each cycle, with the strongest ascent taking place in those portions of the cycle where the force is positive. Furthermore, comparing the two
systems with feedback control, we find that the amount of work needed to transport the particle is larger for the system with $z_0=2\sigma$, than for the one at $z_0=0.32\sigma$. We recall
that the net current is larger at $z_0=2\sigma$, too (see Fig.~\ref{J_period}). Figure~\ref{Work} also shows that the work pertaining to the system under open-loop control has 
qualitatively a similar time dependence, with the numerical values being in between those of the two feedback-control ratchets. 
In other words, in our system feedback control does not necessarily imply that the energy input is smaller 
than that in a comparable open-loop device.
\section{Conclusion}
In this paper we have presented a novel type of a rocking ratchet system, where the particle is subject to a 
space-dependent, asymmetric potential and a time-dependent, homogeneous feedback control force. The control target is the time-delayed mean particle position relative to a switching position, $z_0$. The dynamical properties are mainly studied with a Fokker-Planck equation, where the time-delayed feedback force is introduced {\em ad hoc}. In addition, we have established a connection to a corresponding Langevin equation with mean-field coupling. 

To explore the transport properties of our system we have investigated the net current in dependence of the 
parameters of the control force, that is, delay time, amplitude and switching position. Our results clearly show that
the time delay involved in the feedback protocol is {\em essential} for the creation of a ratchet effect and, thus, for a nonzero net current. A further important ingredient
is the discontinuous dependence of the feedback force on the control target.

An important question for every feedback-controlled system is its efficiency relative to a comparable system under open-loop control. We have found, indeed, that
for a certain range of switching positions (and not too large delay times), the net current is {\em enhanced} relative to the open-loop system. At the same time, however,
the work performed on the particle is larger in the feedback-controlled system. 
This finding is somehow in contrast to a recent result for another ratchet system \cite{sagawa12} where, at the same time, the current was enhanced and 
the work was {\em reduced} by feedback control. Given these subtleties, it seems worth to investigate in more detail
the thermodynamic properties of our model system, including fluctuation theorems and the Jarzynski relation \cite{sagawa12,Seifert}. Another interesting question is to which extent the present feedback scheme, which relies on the (time-delayed) mean particle position as a control target, could be improved to realize, e.g., a larger net current. In fact, as indicated
in Fig.~\ref{J_delay}, the current $J$ does not exceed its value pertaining to the adiabatic limit, at least not for the range of switching positions considered here (recall that this range has been chosen such that the time-averaged force is close to zero).
Therefore, it would be interesting to see whether larger values of $J$ are achievable by choosing another control target (involving, e.g., the force rather than the mean position) or by an otherwise modified control protocol.

So far, there exists no direct experimental realization of the system proposed here, but the main ingredients are already well established.
Indeed, ratchet potentials acting on colloids can be easily realized by using laser beams \cite{lopez08,lee05,hanes12} (optical line trap), and the position of a colloidal particle (or the mean position of many particles) is accessible, e.g., by video microscopy. Moreover, feedback control based on the particle 
position (or mean position) has already been realized experimentally, e.g. in the context of a 
feedback-controlled {\em flashing} ratchet \cite{lopez08} and a Maxwell demon \cite{toyabe10}. Another ingredient, which is indeed crucial in our system, is the presence 
of a time delay. Experimentally, delay arises from various factors \cite{Craig08}, including the time for numerical determination of particle positions via the camera and the time for the decision whether 
to switch the force. In experiments this delay time is about $5-10$ms \cite{lopez08,Craig08} for systems with very few particles and about $60$ms for a large system containing many particles ($N\approx 10^2-10^3$), where larger images are required. 
To judge the impact on transport properties such as $J$, these experimental data for $\tau_\mathrm{D}$ have to be compared
with the intrinsic ("Brownian") time scale $\tau=\sigma^2/D_0$ of a colloidal system. The latter time is
about $1s\lesssim \tau \lesssim 100s$ (for particle sizes of $1\mu m \lesssim \sigma \lesssim 10\mu m$ and diffusion constants
$D_0\approx 10^{-13}$m$^2$/s for colloids in an aqueous solution \cite{dalle11,lopez08,lee05,evstigneev08,tierno10}); therefore, one typically has $\tau>\tau_\mathrm{D}$.
According to the results presented in Fig.~\ref{J_delay}, this is just the regime of ratios $\tau_\mathrm{D}/\tau$, where the current strongly deviates from the adiabatic limit and, in particular, can be larger than in the open-loop protocol. Therefore, we hope that our study will stimulate not only further theoretical work, but also experiments.

\begin{acknowledgments}
This work was supported by the Deutsche Forschungsgemeinschaft through SFB 910 (project B2).
\end{acknowledgments}

\section*{Appendix}
\begin{figure}
\centering
\includegraphics[width=\linewidth]{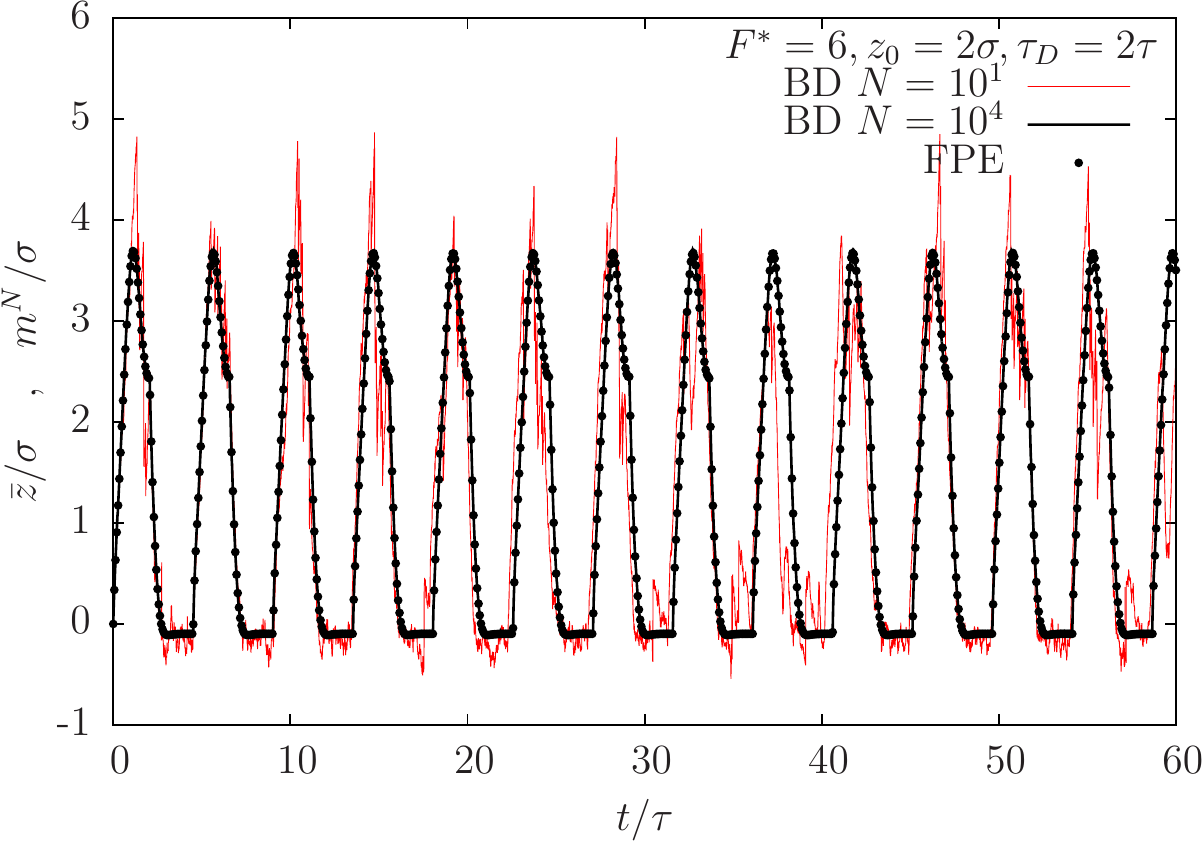}
\caption{(Color online) BD simulation results for the mean particle position as function of time and particle number $N$. Included are FPE results.}
\label{BD}
\end{figure} 
In this Appendix we discuss the connection of the FPE~(\ref{FPE}) and the Langevin equation
%
\begin{gather}
	\label{Langevin}
		\gamma \dot{\chi}_i(t) =  - V'(\chi_i) + F_{\mathrm{fc}}^N(m^N(t-\tau_{\mathrm{D}}))+\sqrt{2\gamma k_\mathrm{B}\mathcal{T}}\xi_i(t)
	\,,
\end{gather}
where $V(\chi_i)$ is given by Eq.~(\ref{Ratchet}), $\xi_i(t)$ represents Gaussian white noise, $i\in \{1,\dots,N\}$, and
\begin{eqnarray} 
\label{f_FC_m}
F_{\mathrm{fc}}^N(t) = -F\cdot \mathrm{sign} \left[ m^N(t-\tau_{\mathrm{D}}) - z_0 \right]
\end{eqnarray}
with 
\begin{equation}
\label{m_def}
m^N(t)=\frac{1}{N}\sum_{i=1}^N \chi_i(t)
\,.
\end{equation}
Thus, $m^N$ is the average of the positions of the $N$ particles.

For the special case $N=1$, one has obviously $m^1(t)=\chi_1(t)$ and thus, $F_{\mathrm{fc}}^{N=1}(t) = -F\cdot \mathrm{sign} \left[\chi_1(t-\tau_{\mathrm{D}}) - z_0 \right]$.
Then, Eq.~(\ref{Langevin}) has the form discussed in earlier studies on delayed Langevin equations, see, e.g., Refs.~\cite{Guillouzic99,Frank03,zeng12}. 
For such systems, the problem in going from the Langevin equation to the FPE is that 
the feedback control force depends on the {\em full} microscopic (stochastic) trajectory
of the particle in phase space up to time t. Therefore, the resulting FPE involves the conditional probability that the particle was at position $z'$ at time $t-\tau_\mathrm{D}$ {\em given} that it is at 
position $z$ at time $t$. An FPE which is formally similar to the usual one [involving only $\rho(z,t)$] can then be obtained by introducing a ``delay-averaged force'', that is,
the integral over space of $F_{\mathrm{fc}}^{N=1}(t)$ times the conditional probability \cite{Guillouzic99,Jiang11}. 

Now we consider the ``mean-field'' limit $N\to\infty$. For each time $t$, averaging over an infinite number of particles is equivalent to averaging over the infinite number of 
realizations of the stochastic force.
Therefore, the quantity $m^N$ in Eq.~(\ref{m_def}) becomes identical to the ensemble-averaged particle position, i.e.,
$\lim_{N\to\infty} m^N(t)=\bar{z}(t)$. As a consequence, the force
$F_{\mathrm{fc}}^N(t)$ does not depend any more on a stochastic quantity, in other words, the information
about the individual stochastic trajectories at time $t-\tau$ is no longer required.
In the ``mean-field'' limit, we can thus consider
the feedback force as a conventional time-dependent force entering the ``mean field" version of Eq.~(\ref{Langevin}), that is,
\begin{align}
	\label{LangevinInf}
	\gamma \dot{\chi}(t) =  - V'(\chi) + F_{\mathrm{fc}}(\bar{z}(t-\tau_{\mathrm{D}}))+\sqrt{2\gamma k_\mathrm{B}\mathcal{T}}\xi(t)
	\,.
\end{align}

From Eq.~(\ref{LangevinInf}), we can derive the FPE in the standard way, i.e., by using the Kramers-Moyal (KM) expansion {\protect\cite{Risken}}. The calculations are, in principle, straightforward;
in particular, there is no problem with multiplicative noise in the mean-field limit. The only uncommon issue 
arises through the fact that our feedback force changes its sign {\em abruptly} when $\bar{z}(t-\tau_{\mathrm{D}})$ crosses $z_0$.  We thus consider in more detail the first (``drift") KM coefficient
\begin{gather}
	D^{(1)}(z,t)=\lim_{\tau\to 0}\frac{1}{\tau}\langle(\chi(t+\tau)-z)_{\chi(t)=z}\rangle
	\,.
	\label{kmcoeff}
\end{gather}
The expression in brackets is evaluated through
\begin{align}
	\chi(t+\tau)-\chi(t)=\int_t^{t+\tau}\mathrm{d}t'\,\dot{\chi}(t')\,
	\label{ztrajdiff}
\end{align}
which can be treated by inserting Eq.~\eqref{LangevinInf} for $\dot{\chi}$ into Eq.~\eqref{ztrajdiff}, iteratively (see \cite{Risken}). Due to the limit $\tau\to 0$ and the noise average incorporated in 
$D^{(1)}$ [see Eq.~(\ref{kmcoeff})]
all terms ${\cal O}(\tau^2)$ as well as terms involving $\langle\xi(t)\rangle$ vanish. The remaining task is to evaluate the term
\begin{align}
	I(t)=\lim_{\tau\to 0}\frac{1}{\tau}\int_t^{t+\tau}\mathrm{d}t'\,F_\mathrm{fc}(\bar{z}(t'-\tau_\mathrm{D}))
	\,.
	\label{coeff_Ffc}
\end{align}
The problem with Eq.~(\ref{coeff_Ffc}) is that, if $F_\mathrm{fc}(\bar{z}(t'-\tau_\mathrm{D}))$ changes its sign in the interval $[t,t+\tau]$, the limit $\tau\to 0$ of $I(t)$ does not exist. Therefore, we make the \emph{assumption} that the time between two switching events has a lower bound, $t^{*}$. Further, we define that at the switching times $t_\mathrm{s}$ (when $\bar{z}(t_\mathrm{s}-\tau_\mathrm{D})\!=\!z_0$) the force $F_\mathrm{fc}(t_\mathrm{s})$ is already set to the new value. 
For all $\tau$ in the interval $\tau\in [0,t^*[$ we then have $F_\mathrm{fc}(t+\tau)=F_\mathrm{fc}(t)$. 
As a consequence, Eq.~\eqref{coeff_Ffc} yields $I(t)=F_\mathrm{fc}(\bar{z}(t-\tau_\mathrm{D}))$ and the first KM coefficient becomes
\label{KMcoeff}
\begin{align}
D^{(1)}(z,t)=\frac{1}{\gamma}(-V'(z)+F_\mathrm{fc}(\bar{z}(t-\tau_\mathrm{D})))\,.
\end{align}
With this expression (and the usual result $D^{(2)}=k_\mathrm{B}\mathcal{T}/\gamma$), one arrives directly at Eq.~(\ref{FPE}).

In order to check our argumentation, we have performed Brownian Dynamics simulations of Eq.~(\ref{Langevin}) for different values of $N$. Representative results for the quantity 
$m^N(t)$ are plotted in Fig.~\ref{BD}, where we have included corresponding results for $\bar{z}(t)$ from the FPE. We see that the results become fully consistent if $N$ is sufficiently large.


\end{document}